\documentclass[aps,prl,twocolumn,superscriptaddress]{revtex4-1}
\usepackage{graphicx}
\usepackage{bm}
\usepackage{color}
\usepackage[latin1]{inputenc} 

\begin{document}
\title{Single-shot initialization of electron spin in a quantum dot using a short optical pulse}

\author{Vivien Loo}
\affiliation{Laboratoire de Photonique et Nanostructures, LPN/CNRS, Route de Nozay, 91460 Marcoussis, France}

\author{Loïc Lanco}
\email[]{loic.lanco@lpn.cnrs.fr}
\affiliation{Laboratoire de Photonique et Nanostructures, LPN/CNRS, Route de Nozay, 91460 Marcoussis, France}
\affiliation{Université Paris Diderot - Paris 7, UFR de Physique, 4 rue Elsa Morante, 75205 Paris CEDEX 13}

\author{Olivier Krebs}
\affiliation{Laboratoire de Photonique et Nanostructures, LPN/CNRS, Route de Nozay, 91460 Marcoussis, France}

\author{Pascale Senellart}
\affiliation{Laboratoire de Photonique et Nanostructures, LPN/CNRS, Route de Nozay, 91460 Marcoussis, France}

\author{Paul Voisin}
\affiliation{Laboratoire de Photonique et Nanostructures, LPN/CNRS, Route de Nozay, 91460 Marcoussis, France}

\date{\today}

\begin{abstract}
We propose a technique to initialize  an electron spin  in a semiconductor quantum dot with a single short optical pulse. It relies on the fast depletion of the initial spin state followed by a preferential, Purcell-accelerated desexcitation towards the desired state thanks to a micropillar cavity. We theoretically discuss the limits on initialization rate and fidelity, and derive the pulse area for optimal initialization. We show that spin initialization is possible using a single optical pulse down to a few tens of picoseconds wide. 
\end{abstract}

\pacs{78.67.Hc,42.50.Pq,03.67.Lx}
\maketitle

Semiconductor quantum dots (QDs) constitute good candidates for the implementation of quantum information processing in the solid state. In the last ten years, the coupling of a single QD transition to an optical cavity mode has allowed numerous cavity quantum electrodynamics realizations, both in the weak and strong coupling regimes \cite{Santori2002,Dousse2010,Reithmaier2004, Yoshie2004, Peter2005}. To benefit from the microsecond spin coherence time \cite{Greilich2006}, for an electron trapped in a quantum dot, spin qubit initialization \cite{Atature2006,Emary2007,Xu2007}, manipulation \cite{Kim2010,Press2008} and readout \cite{Berezovsky2006,Atature2007} are also being widely explored. At the interface, several proposals have focused on the use of light-matter coupling to achieve remote qubit entanglement \cite{Meier2004,Hu2009}. Here the potential of cavity-QD systems for short-pulse spin initialization is explored.

High-fidelity electron spin state preparation has been obtained by several groups worldwide, using different configurations: magnetic field oriented parallel \cite{Atature2006} or perpendicular \cite{Emary2007,Xu2007} to the optical beam for electron spin initialization. Only the perpendicular (Voigt) geometry allows initialization times shorter that the typical spin coherence time: as shown by Emary \textit{et al} \cite{Emary2007}, an initialization fidelity of $99\%$ can be reached in approximately 5 nanoseconds in such a configuration. In this Letter we show that spin initialization can be performed with a single optical pulse down to a few tens of picoseconds wide, using a QD coupled to a high quality factor ($Q$) optical cavity mode, such as the QD-micropillar system sketched in Fig. 1a.

\begin{figure}
\includegraphics[width=8.2cm]{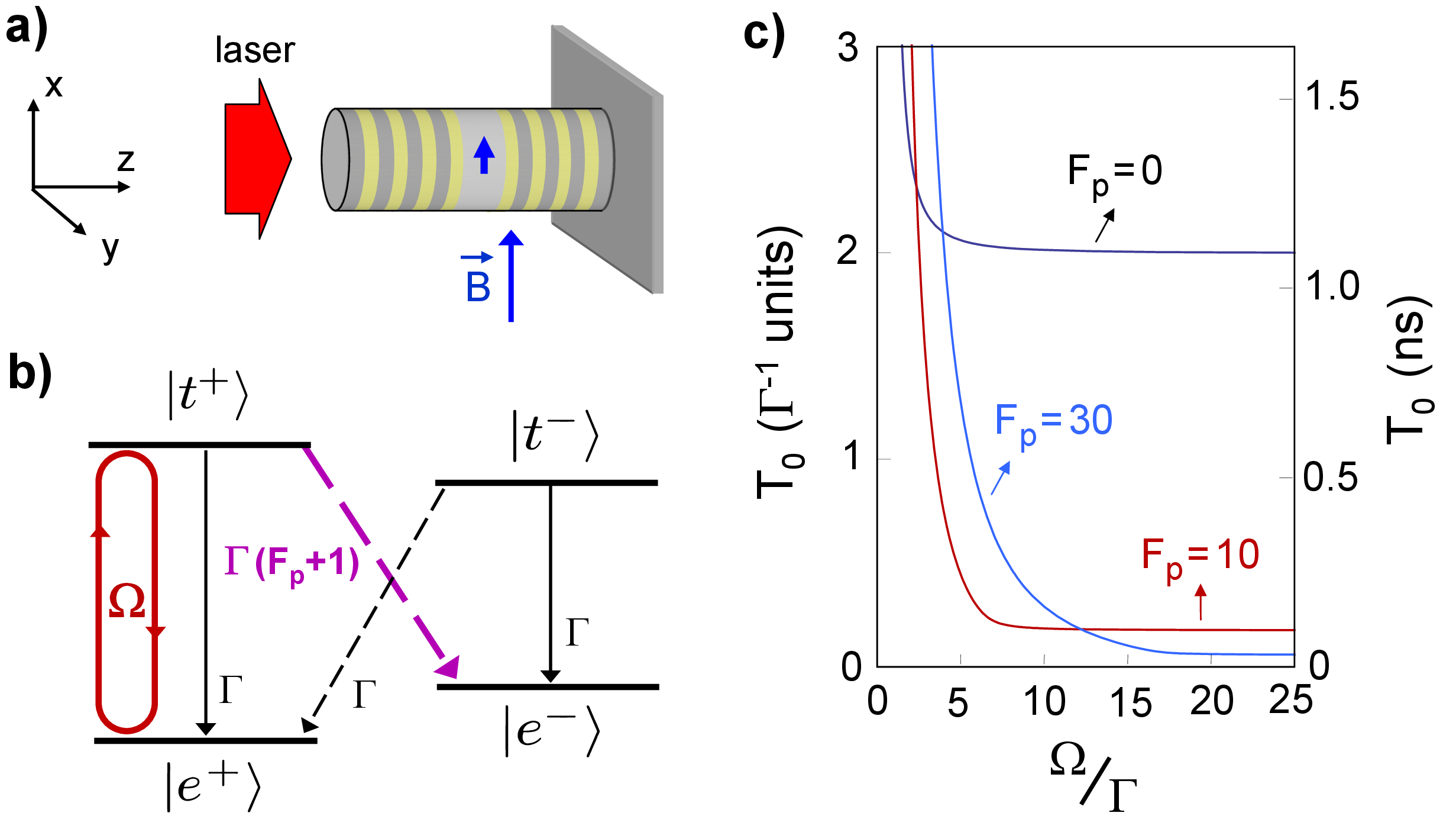}
\caption{\textbf{a)} QD-micropillar device in Voigt geometry.  \textbf{b)} Four-level system and corresponding optical transitions (solid line: X-polarized, dashed line: Y-polarized), with $\Omega$ and $\Gamma$ the pumping and relaxation rates, respectively. \textbf{c)} Initialization time $T_0$ for $F_p$=0, 10 and 30 as a function of $\Omega / \Gamma$.}

\end{figure} 


The corresponding four-level system, in Voigt geometry, is shown in Figure 1b: the low-energy levels are the Zeeman-split electron states with spins along the $x$ axis, that is $\left| e^\pm \right\rangle =\left(\left| \uparrow \right\rangle \pm \left| \downarrow \right\rangle  \right)/\sqrt{2}$, where $\left| \uparrow \right\rangle$ and $\left| \downarrow \right\rangle$ denote the electron spin states quantized along the $z$ axis. The high-energy levels are the Zeeman-split trion states $\left| t^\pm \right\rangle = \left(\left| \uparrow \right\rangle \left| \downarrow \right\rangle - \left| \downarrow \right\rangle \left| \uparrow \right\rangle )( \left| \Uparrow \right\rangle  \pm \left| \Downarrow \right\rangle  \right)/2$, where $\left| \Uparrow \right\rangle$ and $\left| \Downarrow \right\rangle$ denote the heavy-hole spin states in the $z$ direction. In the case of pure heavy-hole states, the $\left| t^+  \right\rangle \rightarrow \left| e^+ \right\rangle$ and $\left| t^-  \right\rangle \rightarrow \left| e^- \right\rangle$ transitions are X-polarized (linearly polarized parallel to the magnetic field), while the $\left| t^+  \right\rangle \rightarrow \left| e^- \right\rangle$ and $\left| t^-  \right\rangle \rightarrow \left| e^+ \right\rangle$ transitions are Y-polarized (perpendicular to the magnetic field) \cite{Emary2007}. We consider a cavity mode in resonance with the $\left| t^+  \right\rangle \rightarrow \left| e^- \right\rangle$ transition, accelerating emission in the mode by a Purcell factor $F_p$.

The initialization process described in Ref. \cite{Emary2007}~ relies on the resonant excitation of the $\left| t^+  \right\rangle \rightarrow \left| e^+ \right\rangle$ transition by an X-polarized laser: the system can then relax either in the desired state $\left| e^- \right\rangle$, or return in $\left| e^+ \right\rangle$ where it can be excited once again. The population remaining in $\left| e^+ \right\rangle$ thus decreases exponentially with a characteristic time $\tau$, given in this case by $\frac{2}{\Gamma}$ where $\Gamma$ is the spontaneous emission rate of the electron-trion transitions. Near-unity fidelity initialization can thus be obtained using optical pulses of durations $\Delta t$ such that $\Delta t \gg \tau$.

Here we show that the selective Purcell acceleration of the $\left| t^+  \right\rangle \rightarrow \left| e^- \right\rangle$ transition, not only decreases the value of this characteristic time down to $\tau=\frac{2}{(F_p+1) \Gamma}$, but also allows to initialize with optical pulses of durations $\Delta t$ shorter than $\tau$. In this regime the initialization process can be schematically viewed as a two-step procedure: complete depletion of $\left| e^+ \right\rangle$ towards $\left| t^+ \right\rangle$ using a pulse of carefully controlled area, followed by a preferential, Purcell-accelerated relaxation towards $\left| e^- \right\rangle$. In the following we analyze the initialization rate and fidelity, and derive the pulse area required for optimal spin initialization. We finally show that micropillar-QD systems with the required properties can be realized using state-of-the-art fabrication techniques.

We consider an X-polarized laser field applied in resonance with $\left| e^+ \right\rangle \rightarrow \left| t^+ \right\rangle$; the second X-polarized transition is detuned by an amount $\Sigma_B$ from the laser field frequency. The laser being non-resonant with the optical mode, the coupling with these transitions occurs mainly through the pillar leaky modes, so that we can assume an equal pumping rate $\Omega$ for both of them. We thus use the hamiltonian given by Emary \emph{et al.} \cite{Emary2007}, except that we define $\Omega$ as the Rabi frequency of the resonantly driven transition, which is twice higher than their definition.

We determine the properties of our system by solving the master equation for the density matrix $\rho$ in the Linblad form ($\hbar=1$ units) \cite{Emary2007}. We assume for simplicity that the Purcell effect only alters the spectrally-matched $\left| t^+  \right\rangle \rightarrow \left| e^- \right\rangle$ transition whose total spontaneous emission rate becomes $(F_p+1) \Gamma$ (that is, $F_p\Gamma$ in the cavity mode, and $\Gamma$ in the micropillar leaky modes), the other decay rates remaining equal to $\Gamma$. We thus use the Lindblad operators of Ref. \cite{Emary2007}, with the same value $\Gamma=1.2\:\mu eV$ ($\Gamma^{-1}=0.55ns$), except for a term $(F_p+1) \Gamma$ instead of $\Gamma$ for the accelerated transition.

We first consider a continous-wave (CW) pumping rate $\Omega$ instantaneously switched on at time $t=0$. In the high-$\Sigma_B$ limit, the varying terms in the hamiltonian oscillate so rapidly that their contribution is averaged to zero, and the density matrix elements $\rho_{ij}(t)$ tend towards constant values $\rho^{(0)}_{ij}$ which are all equal to zero except for $\rho^{(0)}_{e^- e^-}=1$. In the long-time limit, the differences $\rho_{ij}(t)-\rho^{(0)}_{ij}$ decrease proportionnaly to $e^{-\frac{t}{T_0}}$, where the initialization time $T_0$ can be determined analytically. Adapting the calculation of Emary \emph{et al} \cite{Emary2007} to take into account the Purcell acceleration, we find a more general form for $T_0$ as a function of the ratio $r=\frac{\Omega}{\Gamma}$: $T_0=\Gamma^{-1}\left[\alpha-\mu^{1/3}-\beta \mu^{-1/3}\right]^{-1}$, using $\alpha=1+\frac{F_p}{2}$, $\beta=\frac{\alpha^2-r^2}{3}$ and  $\mu=\left(\frac{r}{2}\right)^2+\sqrt{\left(\frac{r}{2}\right)^4-\beta^3}$.

The evolution of $T_0$ is shown in Fig. 1c: as expected, the initialization time is greatly shortened by the Purcell acceleration of the $\left|t^+\right\rangle \rightarrow \left| e^- \right\rangle$ transition. For high enough values of $\Omega$ ($r \gg F_p+1$), an asymptotic regime is reached where $T_0$ tends towards the characteristic time $\tau=\frac{2}{(F_p+1)\Gamma}$; in this regime the initialization speed is not limited by the pumping rate but solely by the desexcitation rate of $\left|t^+\right\rangle$ towards $\left| e^- \right\rangle$. The time required to obtain $99\%$ initialization fidelity is approximately $5T_0$, which is at best $10 \Gamma^{-1}$ in the case $F_p=0$ (no cavity), but decreases to $0.25\Gamma^{-1}$ when $F_p=40$.

If we now reckon with the complete time-dependant hamiltonien (finite $\Sigma_B$), we find that an undesired population will always remain in the state $\left| e^+ \right\rangle$. This arises from the residual Rabi oscillation in the non-resonantly driven, X-polarized transition from $\left| e^- \right\rangle$ to $\left| t^- \right\rangle$. This oscillation, with a generalized Rabi frequency $\sqrt{\Omega^2+\Sigma_B^2}$ and an amplitude proportionnal to $\frac{\Omega}{\sqrt{\Omega^2+\Sigma_B^2}}$, allows a relaxation from $\left| t^- \right\rangle$ to $\left| e^+ \right\rangle$. To determine the corresponding populations, we remark that in the long-time limit the density matrix elements $\rho_{ij}(t)$ still tend towards constants $\rho_{ij}^{(0)}$, except for $\rho_{e^- t^-}$ which tends towards  $\varrho e^{- i \Sigma_b t}$, with $\varrho$ a complex number. Injecting these matrix elements into the master equation leads to the determination of the asymptotic values for the populations $\rho_{ii}^{(0)}$:
\begin{eqnarray}\label{coeffs_rho_zero}
\rho_{e^+ e^+}^{(0)} & = & \left[(F_p+2)^2\Gamma^2+\Omega^2\right]/D \\
\rho_{e^- e^-}^{(0)} & = & 1-\left[(F_p+2)^2\Gamma^2+(F_p+3)\Omega^2\right]/D \\
\rho_{t^+ t^+}^{(0)} & = & \Omega^2/D  \\
\rho_{t^- t^-}^{(0)} & = & (F_p+1) \Omega^2/D 
\end{eqnarray}
with the denominator $D$ given by:
\begin{equation}
D=(8+8F_p+F_p^2)\:\Gamma^2 + 4(F_p+1)\:\Sigma_B^2+2(F_p+2)\:\Omega^2 \nonumber
\end{equation}

The residual population in $\left|e^+\right\rangle$ is an increasing function of the pumping rate $\Omega$; thus, once the latter is set to an appropriate value for fast spin initialization, rising it becomes counterproductive. We find that choosing $\Omega_{opt}=\frac{1}{\tau}=\frac{(F_p+1)\Gamma}{2}$ as the optimal pumping rate provides a good compromise between initialization speed and fidelity. Fig. 2a shows the evolution of $\rho_{e^- e^-}(t)$ for various pumping rates, with $F_p=30$ and $B=6T$ magnetic field, with $\Sigma_B$ determined using the electron and hole transverse $g$ factors $g_{\perp}^{e}=-0.46$ and $g_{\perp}^{h}=-0.29$ \cite{Emary2007}. A pumping rate significantly lower than $\Omega_{opt}$ slows the initialization rate; a higher pumping rate enhances the residual Rabi oscillations in the $\left|e^-\right\rangle \rightarrow \left| t^- \right\rangle$ transition.

\begin{figure}
\includegraphics[width=8.5cm]{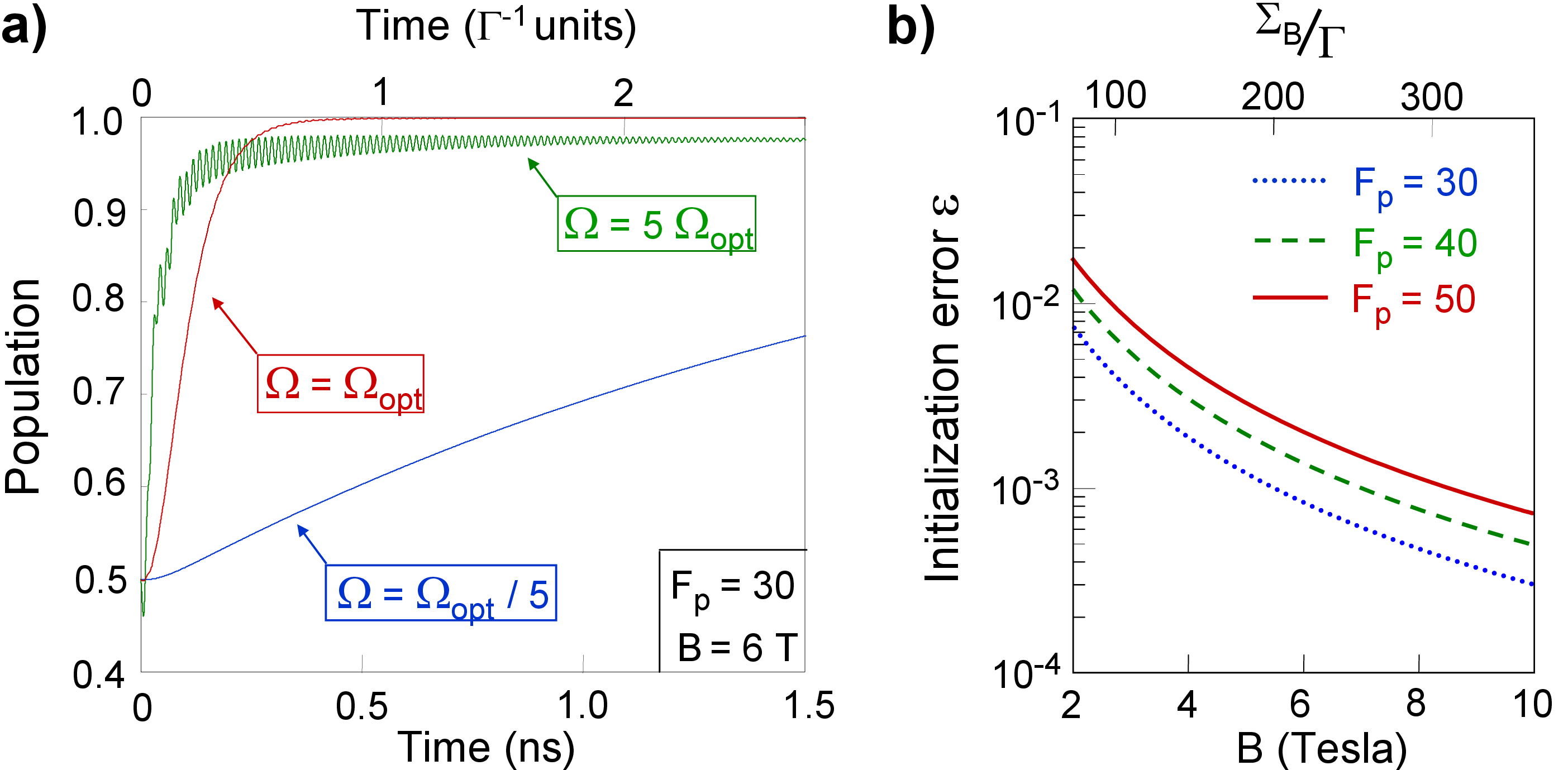}
\caption{\textbf{a)} Population of $\left| e^-  \right\rangle$ as a function of time, numerically integrated for $F_p=30$ and $B=6\:T$, using different pumping rates around the optimal value $\Omega_{opt}=\frac{(F_p+1)\Gamma}{2}$. \textbf{b)} Final initialization error as a function of the magnetic field, for $F_p=30$, $40$ and $50$, for the optimal pumping rate.}

\end{figure} 

%
We now calculate the corresponding initialization error, assuming the CW laser field is turned off at time $t=\Delta t$ with $\Delta t>>\tau$, so that at the end of the pulse the populations have already converged towards their asymptotic values $\rho_{ii}^{(0)}$. At the end of the optical pulse, the population remaining in the state $\left|t^+\right\rangle$ relaxes towards $\left|e^+\right\rangle$ with probability $\frac{1}{F_p+2}$ and towards the desired state $\left|e^-\right\rangle$ with probability $\frac{F_p+1}{F_p+2}$. Similarly, the population in the trion state $\left|t^-\right\rangle$ relaxes towards $\left|e^+\right\rangle$  and $\left|e^-\right\rangle$ with probabilities $\frac{1}{2}$. Thus the initialization error we consider is the residual population in $\left|e^+\right\rangle$ after the pulse, given by $\rho_{e^+ e^+}^{(\infty)}=\rho_{e^+ e^+}^{(0)} + \frac{1}{F_p+2} \rho_{t^+ t^+}^{(0)} + \frac{1}{2} \rho_{t^- t^-}^{(0)}$. We denote $\epsilon$ this final initialization error, which is found to be:
\begin{equation}\label{formula_epsilon}
\epsilon=\frac{2(F_p+2)^2\Gamma^2+\frac{8+5F_p+F_p^2}{F_p+2}\Omega^2}{D}
\end{equation}
In Fig. 2b, $\epsilon$ is plotted as a function of the magnetic field for various Purcell factors, using the optimal pumping rate $\Omega_{opt}$. The error increases when $F_p$ increases, due to the higher value of $\Omega_{opt}$. As expected, the obvious way to limit the initialization error is to increase $\Sigma_b$ in the denominator $D$, and thus the magnetic field.

A major feature of the present initialization procedure, however, is that the practical implementation of spin initialization does not require a pulse duration $\Delta t$ higher than $\tau$. A short optical pulse of carefully controlled area can indeed transfer the population from the $\left| e^+ \right\rangle$ state to the $\left| t^+  \right\rangle$ state; in a second step, the preferential, Purcell-accelerated relaxation of $\left| t^+  \right\rangle$ leads to a nearly complete initialization in $\left| e^- \right\rangle$. For this method to work, a strong Purcell factor is needed, so that the transition back to $\left| e^+  \right\rangle$ is quenched as much as possible. The second issue is that, for the unintended $\left| e^-  \right\rangle \rightarrow \left| t^-  \right\rangle$ transition to remain unaffected by the laser pulse, $\Sigma_B$ must be much higher than the laser spectral width.

To calculate the required pulse area for optimal initialization, we consider the case of a square optical pulse with a constant pumping rate $\Omega$, in the limit $F_p>>1$ and $\Sigma_B$ much greater than both $\frac{1}{\Delta t}$ and $\frac{1}{\tau}$. In such a case, with the laser field switched on at $t=0$, the population of the $\left| e^+ \right\rangle$ state simply evolves as :
\begin{equation}\label{rho_t_square_pulse}
\frac{\rho_{e^+ e^+}(t)}{\rho_{e^+ e^+}(0)}=  \bigg[ \frac{\Omega^2}{2\lambda^2}  +  \Big(1-\frac{\Omega^2}{2\lambda^2} \Big)\cos \lambda t + \frac{ \sin \lambda t }{\lambda \tau}\bigg] e^{-\frac{t}{\tau}} 
\end{equation} 
with $\lambda=\sqrt{\Omega^2-\left(\frac{1}{\tau}\right)^2}$. We remark that $\rho_{e^+ e^+}$ reaches zero at the end of the optical pulse if the latter's duration and pumping rate are related by: 
\begin{equation}\label{eq_square_pulse_duration}
\arctan \lambda \tau + \frac{\lambda \Delta t}{2}= \pi
\end{equation}
The corresponding pulse area, $\theta=\Omega \Delta t$, can be deduced by numerical solving of Eq. (\ref{eq_square_pulse_duration}).

\begin{figure}
\includegraphics[width=8.7cm]{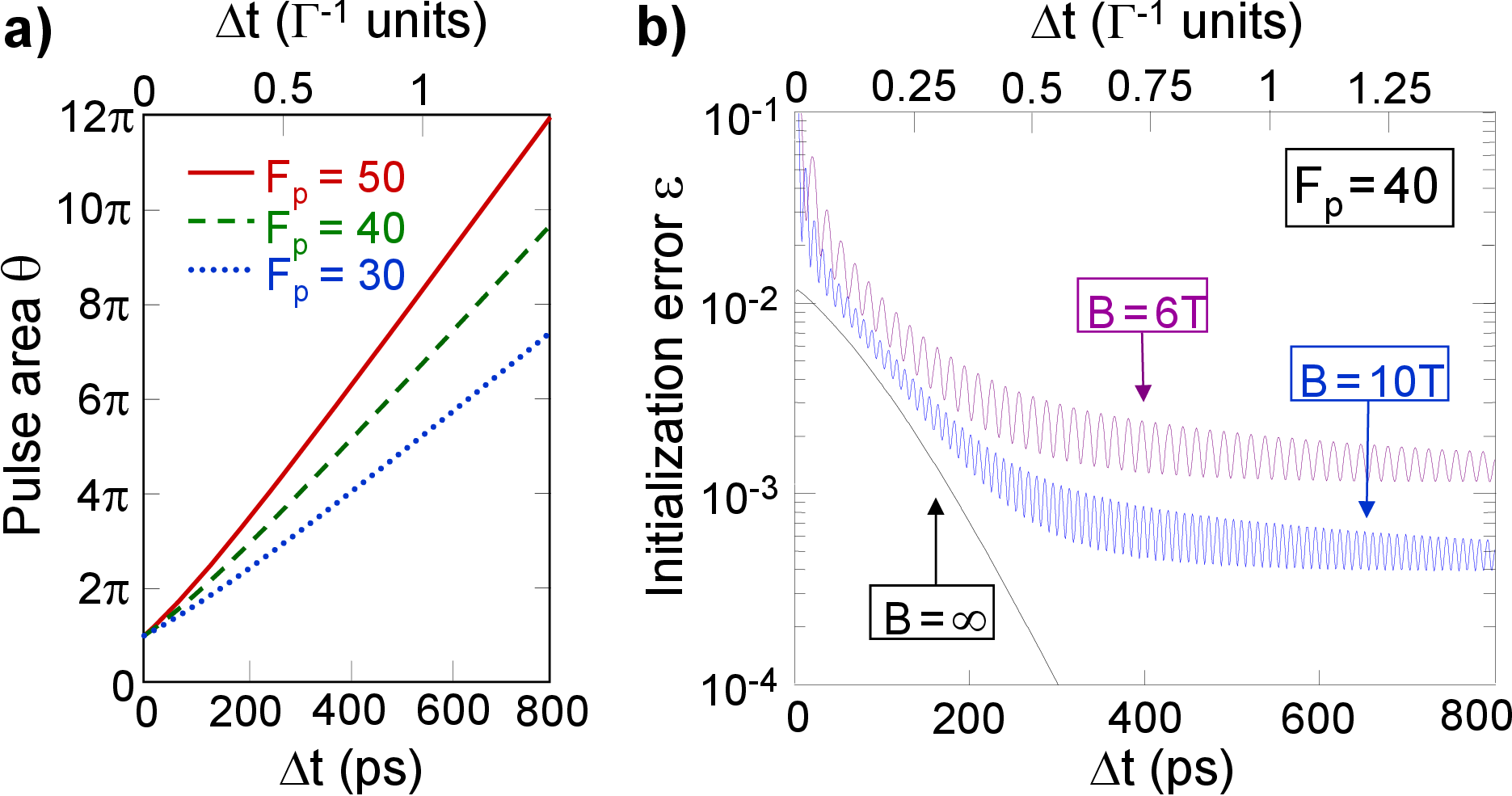}
\caption{\textbf{a)} Pulse area for optimal initialization for $F_p=30$, $40$, and $50$, as a function of the square pulse duration, as deduced from Eq. (\ref{eq_square_pulse_duration}). \textbf{b)} Initialization error corresponding to this pulse area, for $B=6T$, $B=10T$, and in the limit of infinite magnetic field, with a Purcell Factor $F_p=40$.}
\end{figure} 
%

Fig. 3a shows this pulse area for various Purcell factors, as a function of the pulse duration. The limit $\Delta t << \tau$ leads to a pulse area $\theta=\pi$ for complete depletion: this is a well-known property of two-level systems, valid as long as the pulse duration is negligible with respect to the relaxation times. For longer pulses, however, the required area increases with $\Delta t$. When $\Delta t>> \tau$ an asymptotic  behaviour is obtained where $\theta \approx \frac{\Delta t}{\tau}$. This corresponds to $\Omega \approx \frac{1}{\tau}$, which is the above-mentionned optimal pumping rate $\Omega_{opt}=\frac{(F_p+1)\Gamma}{2}$. 

If this optimal pulse area is used, the limit $F_p>>1$ ensures that the population excited in $\left| t^+ \right\rangle$ relaxes towards $\left| e^+ \right\rangle$ only, leading to a perfect spin initialization. In practice, however, the finite values of the Purcell factor and magnetic field lead to a non-zero initialization error, that we investigate numerically. Fig. 3b plots the initialization error as a function of pulse duration, for $6T$, $10T$, and infinite magnetic fields, using $F_p=40$. As expected, for pulse durations $\Delta t>> \tau$ the errors converge towards the values described in Fig. 2b, as given by Eq. (\ref{formula_epsilon}) for $\Omega = \frac{1}{\tau}=\frac{(F_p+1) \Gamma}{2}$. The oscillatory behaviour arises from the non-adiabatic switching of the laser field at $t=0$ and $t=\Delta t$: as shown in Fig. 2a, this leads to residual Rabi oscillations in the $\left| e^-  \right\rangle \rightarrow \left| t^-  \right\rangle$ transition. In the limit of a very short pulse and for an infinite magnetic field, the initialization error tends towards:
\begin{equation}  \label{eq_err_short_pulse_limit}
\epsilon_{\Delta t \rightarrow 0, B\rightarrow \infty}=\rho_{e^+e^+}(0) \: \frac{1}{F_p+2},
\end{equation}
which is the population transferred from $\left| e^+  \right\rangle$ to $\left| t^+  \right\rangle$ multiplied by the probability of the unwanted relaxation from $\left| t^+  \right\rangle$ back to $\left| e^+  \right\rangle$. However, in this regime, and for realistic magnetic fields, the condition $\Sigma_B \Delta t>>1$ is no longer satisfied: the high spectral width of the pumping laser precludes any efficient initialization, as the $\left| e^-  \right\rangle \rightarrow \left| t^-  \right\rangle$ transition is strongly excited.

For these short optical pulses, the approximation of a square pulse is often inappropriate; yet the validity of the method does not depend on the pulse shape as long as the pulse area $\theta=\int{\Omega(t)\mathrm{d}t}$ is carefully controlled. Fig. 4a shows the numerical calculation of the $\left| e^+ \right\rangle$ and $\left| e^- \right\rangle$ populations for a $6\:T$ magnetic field and a $F_p=40$ Purcell factor, using a gaussian pulse with a $40\:ps$ ($0.07 \Gamma^{-1}$) full-width at half-maximum (FWHM). In such a case the magnetic field is high enough to ensure that the unintended $\left| e^-  \right\rangle$ to $\left| t^-  \right\rangle$ transition is non-resonant with the laser pulse, which results in the absence of depletion of the $\left| e^- \right\rangle$ state during the pulse. The preferential relaxation towards $\left| e^- \right\rangle$ leads to a final initialization fidelity around $99\%$, measured at time $t=140ps$ ($0.25 \Gamma^{-1}$); the initialization process is thus approximately 40 times faster than without a cavity. To achieve such a fidelity with a $20\:ps$ FWHM pulse, a Purcell factor $F_p=50$ and a magnetic field $ B=10\:T$ are required. 

\begin{figure}
\includegraphics[width=8.5cm]{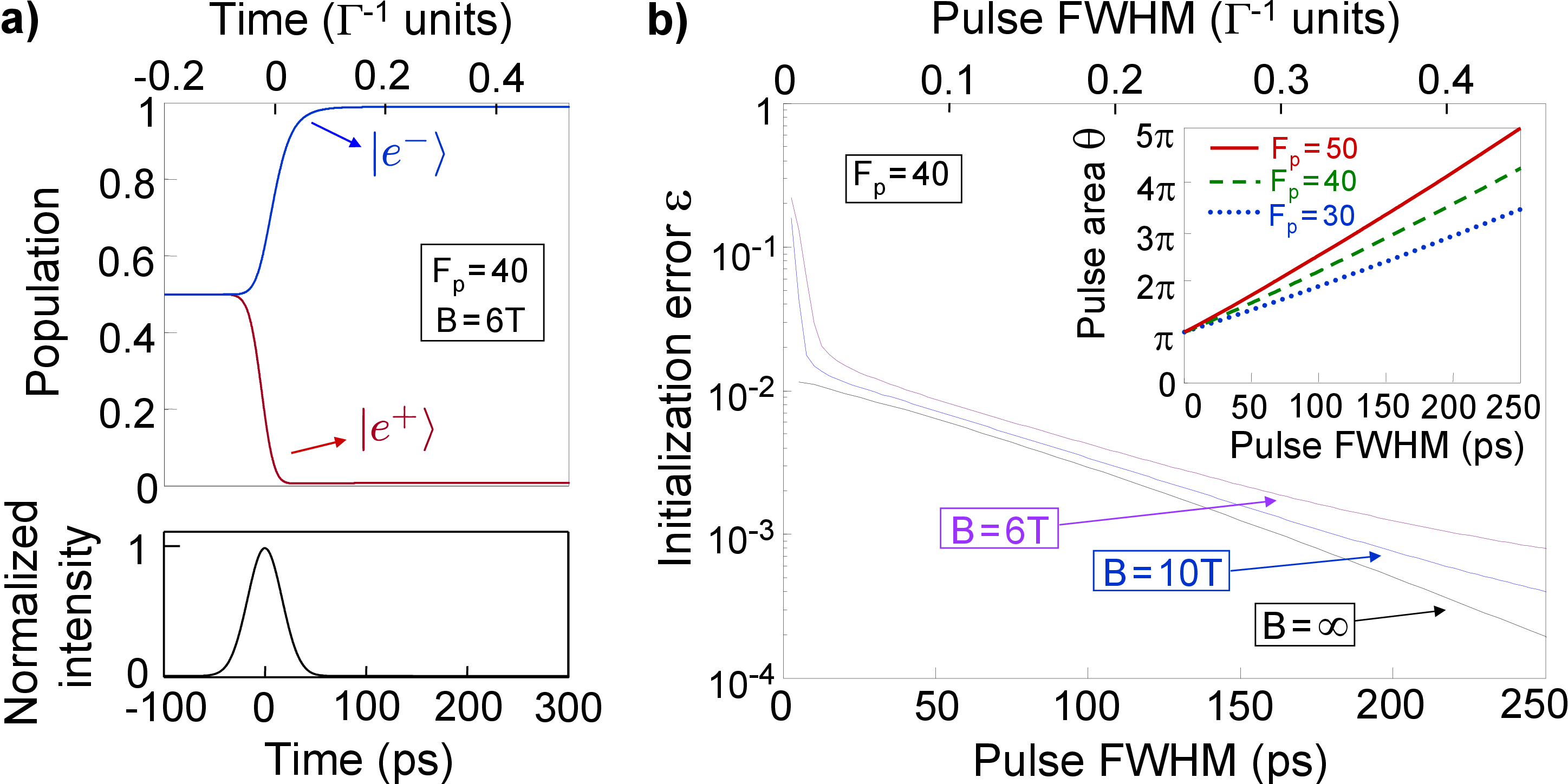}
\caption{\textbf{a)} Optical pulse normalized intensity and corresponding populations of $\left| e+ \right\rangle$ and $\left| e- \right\rangle$ states, as a function of time ($40\:ps$ FWHM gaussian pulse, with $F_p=40$ and $B=6T$). \textbf{b)} Initialization error as a function of the gaussian pulse width for $6\:T$, $10\:T$, and infinite magnetic fields, with $F_p=40$. Insert: required pulse area for optimal initialization using a gaussian pulse, for $F_p=30$, $40$, and $50$.}
\end{figure} 


Fig. 4b shows the expected initialization error for $6T$, $10T$, and infinite magnetic fields, using $F_p=40$. As expected, the initialization error in the limit of infinite magnetic field and very short pulse duration remains given by Eq. \ref{eq_err_short_pulse_limit}. The main difference with Fig. 3b is that the smooth evolution of $\Omega(t)$ allows a monotonic decrease of the initialization error when the pulse width increases. As long as the pulse FWHM is larger than a few tens of picoseconds, thanks to the limited spectral width of the gaussian pulse, the finite value of the magnetic field does not significantly alter the initialization efficiency.

The insert of Fig. 4b summarizes the required pulse area, as a function of the pulse FWHM, displaying the same behavior than in Fig. 3a for square pulses: the pulse area for optimal initialization grows above $\pi$ when the pulse width increases. We note that the pulse area shows no significant dependence on the magnetic field in the $6T-10T$ range, and that no fine-tuning is required for practical implementation: for example, a $2\%$ shift of the pulse area will result in an increase of the initialization error of the order of a few $10^{-3}$. 

We now show that this spin initialization procedure can be implemented in realistic QD-cavity systems. The current state-of-the-art of nano-fabrication techniques allows to realize, in a both deterministic and scalable way, QD-micropillar systems whose cavity mode is spatially and spectrally matched with a selected QD transition \cite{Dousse2008}. Micron-size pillars can be realized with quality factors up to $10^5$ \cite{Reitzenstein2007}. For example, a $3µm$ diameter AlAs/GaAs micropillar (effective mode volume $V \approx 1~µm^3$ and effective index $n \approx 3.4$) with a quality factor $Q=2.5\ 10^4$, embedding a single InAs/GaAs QD with $f=10$ oscillator strength (transition wavelenth $\lambda = 0.94µm$) will present a Purcell Factor $F_p \approx 40$ while staying in the weak-coupling regime \cite{Reithmaier2004}.

A key condition to the validity of this initialization procedure is that the undesired relaxation from $\left| t^+  \right\rangle$ back to  $\left| e^+ \right\rangle$ does not experience any significant Purcell acceleration. Using the transverse $g$ factors of Ref. \cite{Emary2007}, we find that a $6T$ magnetic field corresponds to a $160 \: µeV$ detuning between the $\left| t^+  \right\rangle \rightarrow \left| e^+ \right\rangle$ transition and the cavity mode, whereas the mode resonance half-width is $25µeV$ for $Q=2.5\ 10^4$: this corresponds to a Purcell factor $F_p'\approx 1$ ($F_p'\approx 0.3$ for $B=10$~T). The small Purcell effect experienced by the Y-polarized $\left| t^-  \right\rangle \rightarrow \left| e^+ \right\rangle$ transition (the Zeeman splitting between the Y-polarized transitions is only $60µeV$ for $B=6$~T) does not alter the validity of the method.

We also point out that the presence of heavy/light-hole mixing in the QD can affect the trion states, resulting in a rotation of the polarization axis in the $x-y$ plane \cite{Xu2007}; furthermore, a small alteration of the beam polarization may also occur in the Bragg mirrors, due to the residual ellipticity of the micropillar (ellipticities of the order of $5.10^{-4}$ are the current state-of-the-art \cite{Reitzenstein2007}). For simplicity the orthogonal polarization states of the optical transitions have been called $X$ and $Y$, whether or not they correspond to linear polarizations along the $x$ and $y$ axes of Fig. 1a. For efficient spin initialization the laser polarization has to be adjusted to ensure negligible excitation of the Y-polarized transitions.

Having derived the optical pulse area required for optimal initialization, as well as the corresponding fidelity and experimental conditions, we conclude that spin initialization can be implemented using short optical pulses down to a few tens of picoseconds wide. This work is partially supported by ANR project n°ANR-JCJC-09-0136.

\end{document}